\documentstyle[11pt]{article}
\catcode`\@=11
\begin{document}\openup6pt

\title{ ACCELERATING  UNIVERSE IN  MODIFIED THEORIES OF  GRAVITY }

\author{ B. C. Paul\thanks{bcpaul@iucaa.ernet.in}, P.S. Debnath\thanks{parthasarathi6@hotmail.com} and S. Ghose \\
         Physics Department, P. O. North Bengal University \\
       Siliguri, Dist. : Darjeeling, PIN : 734 013, India}

\date{}

\maketitle

\begin{abstract}
We  study cosmologies in  modified theories of   gravity considering Lagrangian density $f(R)$ which is a polynomial function of  scalar curvature ($R$) in the Einstein-Hilbert action in vacuum. The field equation obtained from the modified action corresponding to a Robertson-Walker metric is highly  non-linear and not simple enough to obtain analytic solution. Consequently we adopt a numerical technique to study the evolution of the FRW universe.
  A number of evolutionary phases of the universe including the present accelerating phase  are found to exist in the higher derivative theories of gravity. The cosmological solutions obtained here are new and interesting. We study modified theory of gravity as a toy model to explore the past, the present and predict the future evolution. It is found that all the models analyzed here can reproduce the current accelerating phase of expansion of the universe. The duration of the present accelerating phase is found to depend on the coupling constants of the gravitational action.  The physical importance of the coupling parameters those considered in the action are also discussed.\\

PACS numbers : 98.80.Cq
\end{abstract}
\vspace{2.5in}
\pagebreak

$\bf I.\ \ INTRODUCTION :$

The last couple of years witnessed  a modest progress in our understanding of the observed universe because of the advent of new cosmological precision tests, capable of providing physics at very large redshifts. The luminosity curve of type Ia supernovae [1], the large scale structure [2] and  the anisotropy of the CMB [3] favour a spatially flat universe. Recent decade is witnessing a paradigm shift in cosmology  from speculative  to experimental science due to a large number of observational inputs. The has been predicted that the present universe is passing through a phase of the cosmic acceleration. It is also believed that the universe might have emerged from an inflationary phase in the past. A large number of cosmological models were proposed in Einstein's gravity with early inflationary scenario in the last three decades which works well. However, the recent prediction that  the present universe is passing through an accelerated phase of expansion is interesting and a proper cause is yet to be understood. It is thought that the cause of the present acceleration of the universe might be due to dark energy in the universe. However, the concept of dark energy in the  Einstein's gravity with normal matter or fields cannot be implemented. Consequently it is a  challenging job in the theoretical physics to frame a theory for cosmological evolution  which could address the origin 
 of dark energy also. It is known from the cosmological observations that the dark energy content of the universe is about $70 \%$ to that of the total  energy budget of the universe. As mentioned earlier the usual fields available in the standard model of the particle physics are  not enough to account for the huge dark energy reservoir in the universe, a modification to the Einstein's field equation either in the gravitational or in the matter sector, therefore, is essential to  accommodate  the present cosmological observations. The issue of dark energy has been taken up in a gravitational theory in presence of a cosmological constant [4, 5]. However, the vacuum energy density in such a theory remains constant in the course of cosmic evolution and it is also true that there are known contributions into vacuum energy which are several orders of magnitude greater than the allowed cosmological values. These observations led us to look for an alternative model or a new physics [6-8]. It has been proposed recently that gravity itself, if properly modified, could account for the recent cosmic acceleration [9, 10]. The standard Einstein's gravity may be modified at low curvature by including  the terms those are important precisely  at low curvature. The simplest 
possibility is to consider a $\frac{1}{R}$-term in the Einstein's-Hilbert action (it may originate from M-theory) [11].
 Carroll $\it et\;al.$ [12] also  suggested that such a theory may be suitable to derive cosmological models with late accelerating phase.  Although a theory with $\frac{1}{R}$-term in the Einstein's gravity accounts satisfactorily the present acceleration of the universe, it is realized that inclusion of such terms in the Einstein-Hilbert action leads to instabilities [13].
   Subsequently, it has been shown that further addition of a  $R^2$- 
 term [14] or $ln(R)$ term [15]  to the Einstein's gravitational action leads to a consistent modified theory of gravity which may pass 
 satisfactorily solar system tests, and  free from instability problem. It is known that the modified gravity 
 with a positive power of the curvature scalar (namely, $R^2$-term) [15-18] in the Einstein-Hilbert action admits early inflation.
The modified gravity with negative powers  of the curvature in the Einstein-Hilbert action is recently becoming popular as it might effectively behave as a dark energy candidate. Consequently the theory might satisfactorily describe the recent cosmic acceleration [9-11]. So it is reasonable to explore a theory which could accommodate an inflationary scenario of the early universe and an accelerating phase of expansion at late time followed by a matter dominated phase. As a result modified theory of gravity which contains  both positive and negative powers of the curvature scalar ($R$)
namely, $f(R) = R+\;\alpha R^m+\;\beta \frac{1}{R^n}$ where $\alpha$ and $\beta$ represent coupling constants with arbitrary constants  $m$ and $n$ are considered  for exploring cosmological models.  It is known that the term $R^m$ dominates and it permits power law inflation if 
 $1 < m \leq 2$, in the large curvature limit.  It may be mentioned here that  an inflationary scenario driven by vacuum trace anomaly which corresponds to 
 $m=2$ and $ \beta=0$ was first obtained by Starobinsky [17] for describing early inflation. Recently, in the low curvature limit,  a number of  $f(R)$ models have been proposed in order to accommodate   universe with late acceleration using a modified gravity namely, $f(R) = R- \frac{\lambda}{R^n}$ with $n > 0$ [11, 12]. In the metric approach, it was shown that the model is not suitable because it does not  permit a matter era [19]. 
Recently, it is known that modified gravity namely, $f(R) = R + \alpha R^m \;$ is also  not cosmologically viable because it does not permit a consistent scenario accommodating a matter dominated era at  late time. It is found that instead of matter era, one ends up with a radiation era ( $a(t) \sim \sqrt{t}$ ). On the other hand, $R^m$ model permits matter dominated universe ($a(t) \sim t^{2/3}$)  but  it fails to connect to a late accelerating phase. In {\it Ref.} [19], it was shown that the models of the type where Lagrangian density, $f(R) = R - \frac{\lambda}{R^n}$ with $n > 0$ and $f(R) = \alpha  R^m$ with $m \neq 1$ are not viable for a realistic cosmological scenario as they do not permit matter epoch although late acceleration can be realized [20].  Recently,  modified gravity with power law  in $R$, i.e.,   $f(R)$ -gravity is examined and found that a large class of models including $R^m$ model does not permit matter dominated universe. Capozziello {\it et al.} [21] criticized the claim made in {\it Ref.}  [19]. Tsujikawa [22] derived observational signature of $f(R)$ dark energy models that satisfy cosmological and local gravity constraints fairly well. The modified $f(R)$-gravity is found to be consistent with realistic cosmology in some cases [23]. However, no definite physical criteria known so far  to select a particular kind of theory  capable of matching the data at all scales. However, modified gravity namely, $f(R) \sim e^{R}; \; or \; \; \log R \;$ may be useful to build a viable cosmological model as they permit a matter dominated phase before an accelerating phase of expansion. The motivation of the  paper  is to obtain cosmological solutions  considering non-linear terms in $R$ in the Einstein-Hibert action. We explore different phases of expansion of the universe from early era to the present and would like to understand the future evolution in the framework of higher derivative gravity.The corresponding field equation obtained from the above gravitational action is a fourth order differential equation of the scale factor of the universe ( $a(t) $ ). As the field equation is  highly non-linear and not simple enough to obtain an analytic solution we  adopt here a numerical technique to solve it. Our approach here is similar to that adopted earlier in {\it Ref. } [24] which was recently employed by two of us  in {\it Ref.} [25]. In this approach the field equations are first expressed in terms of two functions namely, deceleration parameter ($q$) and Hubble parameter
 ($H$) and its derivatives respectively, which are  then solved numerically.  \\
The plan of the paper is as follows: in sec.II, we set up the relevant field equations in the modified theory of gravity, 
in sec. III, cosmological evolutions are predicted in different models depending upon the coupling  parameters of the action adopting numerical technique. Finally in sec.IV, we summarize the 
results obtained.

$\bf II.\ \ THE \; MODEL :$

We consider a gravitational action with non-linear terms in the scalar 
curvature $(R)$ which is given by
\begin{equation}
{\large I} = - \int  \left[\frac{1}{2} f(R)   + L_{m}  \right] \sqrt{- g} \; 
d^{4}x
\end{equation}
where $8\pi G=c=1$, $g$ is the determinant 
of the four dimensional metric and $R$ is the scalar curvature. Here $f(R)$ 
is a function of $ R$ and its higher power  and $ { L_m}$ represents the matter Lagrangian. Variation of the action (1)
with respect to the metric yields 
\begin{equation}
G_{\mu\nu} = R_{\mu\nu} -\frac{1}{2}R g_{\mu\nu} = T_{\mu\nu}^c + T_{\mu\nu}^M,
\end{equation}
where $T_{\mu\nu}^M$ represents the contribution from matter fields scaled by a factor of $\frac{1}{f'(R)}$
 and $T_{\mu\nu}^c$ denotes the contribution that originates from the curvature to the effective stress energy tensor. Here, $T_{\mu\nu}^c$ is actually
 given by
\begin{equation}
T_{\mu\nu}^c = \frac{1}{f^{'}(R)}\left[\frac{1}{2}g_{\mu\nu}(f(R)-Rf^{'}(R))+
f^{'}(R)^{;\alpha\beta}(g_{\mu\alpha}g_{\nu\beta}-g_{\mu\nu}g_{\alpha\beta})\right],
\end{equation}
where prime represents the derivative with respect to the Ricci scalar ($R$). We are interested to study the role of the geometry alone in driving cosmological evolution in this paper, so we set  $L_m =0$ which leads to $T_{\mu\nu}^M =0$ in the subsequent sections. 
The flat Robertson-Walker spacetime is given by 
\begin{equation}
ds^{2}=dt^{2}-a^2(t)\left[dr^{2}+r^{2}(d{\theta}^{2}+sin^{2}{\theta } \; 
d{\phi}^{2})\right ]
\end{equation}
where $a(t)$ is the scale factor of the universe. Using the metric (4) in  the field eqs. (2)  (see also Ref. [24]) we obtain   
\begin{equation}
3\frac{\dot a^2}{a^2}= \frac{1}{f^{'}}\left[\frac{1}{2}(f-Rf^{'})-3\frac{\dot a}{a}\dot{R}f^{''}\right],
\end{equation}
\begin{equation}
2\frac{\ddot a}{a} + \frac{\dot a^2}{a^2} = -\frac{1}{f^{'}}\left[2\frac{\dot a}{a}\dot{R}f^{''} + 
\ddot{R}f^{''} +\dot{R^2}f^{'''} -\frac{1}{2}(f-Rf^{'})\right],
\end{equation}
where an over dot indicates derivative with respect to the cosmic time $t$. 
The scalar curvature is given by
\begin{equation}
R= - \; 6  \; \left( \frac{\ddot{a}}{a} + \frac{\dot a^2}{a^2}\right).
\end{equation}

The Ricci scalar $R$ involves second order time derivative of the scale factor $a$. As the  eq. (6) contains $\ddot{R}$ terms, one 
actually has a system of fourth order differential equations of scale factor. In the next sections we consider modified gravity with  $f(R)$ of the forms :  (A) $f(R) = R +\alpha R^2 -\frac{\mu^4}{R}$ ,  (B) $f(R) = R + b\; ln(R)$ and  (C)
$f(R)=R + m\; e^{[- nR]}$ to explore cosmic evolution. In the above,  $\mu $ , $\alpha$ , $b$ , $m$ and $n$ are constants  and  $\mu$ has a dimension  
 of $R^{\frac{1}{2}}$  [26] i.e.
 that of (time)$^{-1}$, $\alpha$ has a dimension of $R^{-1}$ i.e. (time)$^{2}$. \\

${\bf III.\; COSMOLOGICAL\; SOLUTIONS:}$

Using eqs. (5) and (6) we obtain
\begin{equation}
\dot{H} = \frac{1}{2 f^{'}} \left[(H\dot{R}-\ddot{R})f^{''}- \dot{R}^2 f^{'''} \right],
\end{equation}
where $H =\frac{\dot a }{a}$ is the Hubble parameter. As both $R$ and $H $ are functions of $a$ and its derivatives, 
eq. (8) is highly non-linear and a differential equation of fourth order in scale factor ($a(t)$). It is not simple to determine  analytic solution of  the scale factor of the universe with cosmic time in closed functional form. Hence adopt numerical technique to study the behaviour of the cosmological models based on the parameters of the modified gravitational action. For simplicity we consider $f(R)$ of three different forms in the  next sections. 
 
$\bf{Case\; A :}$

In this case we consider modified  gravitational action where $f(R)$ is given by
\begin{equation}
f(R) = R +\alpha R^2 -\frac{\mu^4}{R}.
\end{equation}
The corresponding field eq. (8) becomes
\begin{equation}
\dot H =\frac{1}{1+ 2\alpha R + \frac{\mu^4}{R^2}}\left[\frac{\mu^4}{R^2}\left( \frac{\ddot R}{R}-\frac{H\dot R}{R}-3\frac{\dot R^2}{R^2} \right)+\alpha(H\dot R -\ddot R )\right].
\end{equation}
 The above equation is highly non-linear,  however, one can obtain asymptotic solutions corresponding to different epoch which are 

(i) an exponential expansion $a(t) \sim e^{ H_o t}$ is permitted when $q = -1$ in the early era,

(ii) an accelerating universe  with $a(t) \sim  t^2$ at a later epoch admitting $q= - \frac{1}{2}$. The variation of $q$ with $a(t)$ in linear evolutionay and exponential phase are shown in figs.1 and 2 respectively. 

\input{epsf}
\begin{figure}
\epsffile{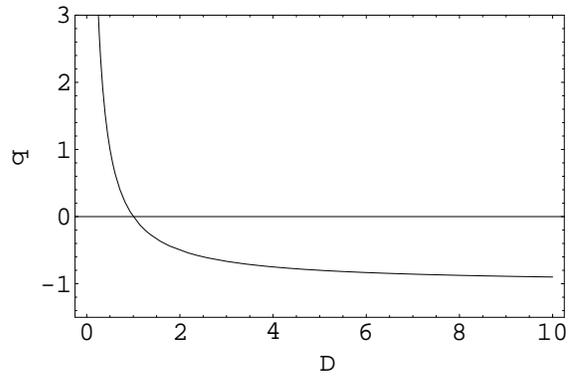}
\caption{  shows the plot of  $q$  Vs  $a(t)$ in the unit of $ln (t)$.}
\end{figure} 
\input{epsf}
\begin{figure}
\epsffile{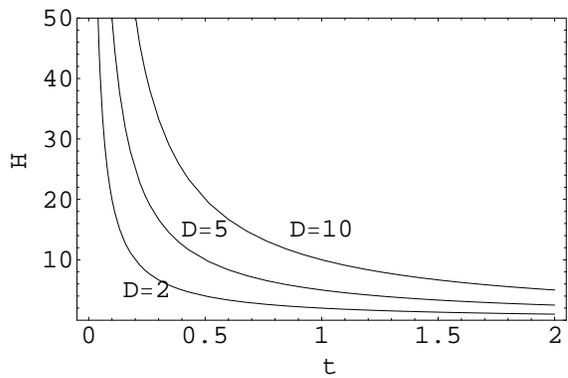}
\caption{  shows the plot of $H$ with $t$ for different $D$.}
\end{figure}

Now using  deceleration parameter 
$(q)$      
\begin{equation}
q =-\frac{\ddot{a}a}{\dot a^2} =-\frac{\dot H}{H^2}-1, 
\end{equation}
 one can study the evolution of the universe numerically. Since, $q $ is a function of $H$ and its derivative, we can re-write the eq. (10) 
in terms of a second order differential equation in $q$ and $H$ to begin with. Since $q$ contains terms with
  $ \ddot a$, one can replace terms with fourth order derivative of the scale factor in eq. (10) by $\ddot q [H]$. The 
   functions $q$ and $H$ are, however, not independent. The time derivatives in the above equations may now be replaced 
   by the derivatives with respect to $H$ using eq. (11). We get the following non-linear differential equation
\begin{equation}
q^{''} + u(q,H) {q'}^{2} + v(q,H) q' + w(q,H) = 0
\end{equation}   
 where $$u(q,H) = -\frac{(2q+4)\mu^4 +216 \alpha (q-1)^4 H^6}{(q^2-1)[\mu^4 -216\alpha(q-1)^3 H^6]}, $$
   $$v(q,H) = -\frac{(4q+7)\mu^4 +216 \alpha (8q+5)(q-1)^3 H^6}{(q+1)H[\mu^4 -216\alpha(q-1)^3 H^6]},$$
  $$ w(q,H) = \frac{(q-1)[3\mu^4(2q+1) +1296 \alpha (q+1)(q-1)^3 H^6 - 36(q-1)^2 H^4]}
    {(q+1)H^2[\mu^4 -216\alpha(q-1)^3 H^6]},$$
where the prime indicates a differentiation w.r.t. $H$ and the functions $u,\;v$ and $w$ depend on  $\alpha$ and $\mu^4$. The above equation, although highly nonlinear, is a second order differential equation in $q$.
  Here both $q$ and $H$ are time dependent and cannot be solved exactly to obtain a known functional form. We solve it numerically in the next section following the approach adopted in Ref. [24]. As $\frac{1}{H}$ is a measure 
of the age of the universe and $H$ is a monotonically decreasing function of the cosmic time, eq. (12) may be 
used to study qualitatively  the evolution  of the universe in terms of  $q$. Since eq. (12) is a second order
 differential equation, to solve it numerically we assume two initial conditions (here, $q[H]$
  and $q'[H]$), for a given value of $H$.   We choose units so that $H_o$, the present value of $H$, is unity and pick up 
sets of values of $q$ and $q'$ for $H =1$ (i.e. the present values) from the observationally consistent region [5]. We 
  plot $q$ with  $H$  for different configuration of the system. As the inverse of $H$ gives us an estimate for the
   cosmic age, future evolution is understood from the region $H < 1$ and the past from $H >1$ in the ($q$-$H$)-plane.
 Since the present universe is accelerating, we use a negative $q$ at the present epoch, $H=H_0=1$ and the universe 
  has entered into this $q$ nagative phase (i,e, acceleration) of expansion, only in the recent past.  The model may be
   used to predict also the future course of the evolution of the universe. We note the following :
      
(i)   For a given value of $\alpha$ and $\mu$, the variation of $q$ with $H$ is shown in fig. 3. We choose $\alpha = 2$, $\mu^4 = 12$. In the graph time increases from right to left along the horizontal $H$-axis. The upper half of the $H$-axis represents decelerating phase and the lower half represents the accelerating phase of the universe.  It is evidentthat the universe entered into the present accelerating phase in the recent past, the rate of acceleration will increase further which will attain a maximum thereafter it decreases. There will be slowing down of the cosmic acceleration leading to an epoch when the universe expands without acceleration (which is transient and occures at $H=0.375$) followed by another phase of expansion. In this case the  expansion of the universe  will be accelerating once again. The universe transits from decelerating to accelerating phase at $H=1.36$. We note that, in this case, the universe remains in the accelerating phase once it transits from the decelerating phase.

\input{epsf}
\begin{figure}
\epsffile{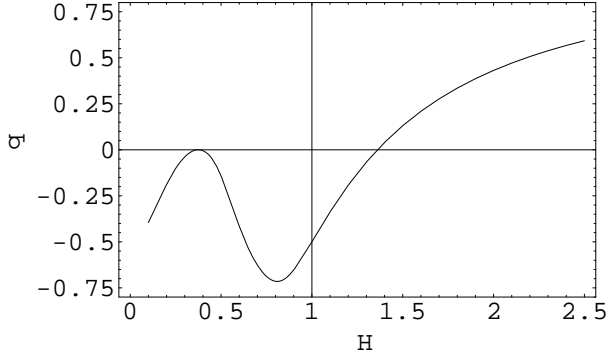}
\caption{  shows the plot of $q$ with $H$ for $\mu^4=12$ and $\alpha= 2$ . Here we choose the initial conditions as 
$q$[1]= -0.5, $q'$[1]=1.655.}
\end{figure} 

(ii)  For a given set of values of $\alpha$ and $\mu$ we  plot $q$ vrs. $H$ for different initial values  of $q'[1]$ 
with $q[1]=- 0.05$, when $\mu^4=12$ and $\alpha=2$, the curves are  shown in fig. 4. It is evident that the  universe 
 entered into the accelerating phase in the recent past followed by another phase of deceleration.   The  duration for which the universe transits from the present accelerating phase depends on the initial values of $q'[1]$,and  the duration increases with increasing  initial values of  $q'[1]$. We note the following (i) for $q'[1]=1$ the universe transits from deceleration to acceleration at $H= 1.43$ and acceleration to deceleration at $H = 0.587$; (ii) as initial value of $q'$ is increased, the critical point shifts toward left, i.e. occurs at a later epoch. 

\input{epsf}
\begin{figure}
\epsffile{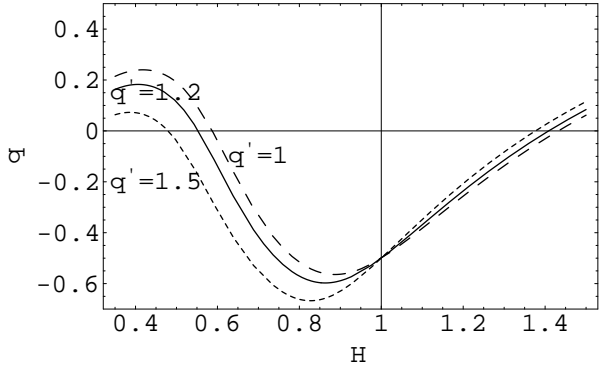}
\caption{ shows the plot of $q$ with $H$ for different initial conditions $q'[1]=1, $ $q'[1]= 1.2 $, 
$q'[1]$= 1.5 with $q[1]$= -0.5.}
\end{figure} 

(iii)  We plot $q$ vs. $H$ for different values of $\alpha$ using $q[1] = - 0.5$, $q'[1] = 1.2$ and $\mu^4=12$ which 
is shown  in fig. 5. The figure shows how the nature of the curves depend on the initial conditions and that there will be another sign change of $q$, in the near future. It is evident that the universe transits from decelerated phase to accelerated phase only in the recent past($H=1.4$). The universe may  transit again to a  decelerating phase followed by another  accelerating phase if $\alpha > 0.815$ otherwise it will be always in the accelerating phase. However, the rate of acceleration depends on the coupling constant $\alpha$. It is evident that as the values of  $\alpha$ is increased the corresponding  duration of present accelerating phase will get reduced, however, the late decelerating phase in future will be enhanced. The universe in  the remote future once again might enter into an accelerating phase as there will be one more sign flip in $q$. The plot for $\alpha = 0.815$ is interesting as it decides whether the universe will have another phase of acceleration or not.

\input{epsf}
\begin{figure}
\epsffile{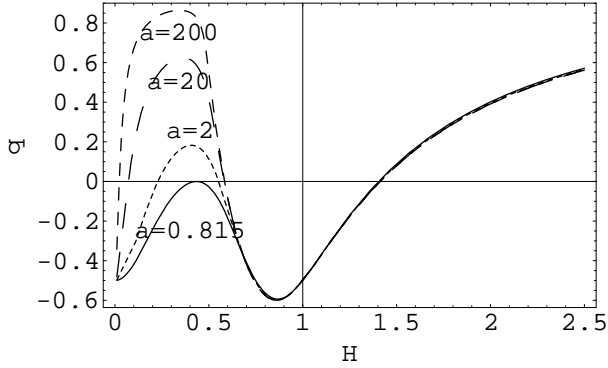}
\caption{ shows the plot of $q$ with $H$ for different value of $\alpha$ ($a$
 represents $\alpha$). Here we take $\mu^{4}=12$ and choose 
the initial conditions as $q$[1]= -0.5,
 $q'$[1]=1.2.}
\end{figure}

(iv) The variation of $q$ with $H$ for different values  of coupling constant $\mu^4$ is shown in  fig. 6
  for $\alpha = 2$. It is evident that if inverse Ricci scalar term in the action is absent, then it permits a  universe  which transits from a decelerating phase to an accelerating phase in the recent past $(H=1.4)$ allowing a further  sign flip in $q$ leading to a transition from  accelerating to decelerating phase only.  However, for $\mu^4 \neq 0$, an interesting 
evolutionary behaviour of a  universe with three sign flips of $q$ leading to a universe from accelerating to decelerating followed by a phase decelerating to accelerating and in future deceleration to acceleration might happen. However, the rate of acceleration changes with  $\mu$. As $\mu^4$ increases, the period of deceleration to another acceleration in recent future will decrease and finally period vanishes at $\mu^4 = 26.46$. 
\input{epsf}
\begin{figure}
\epsffile{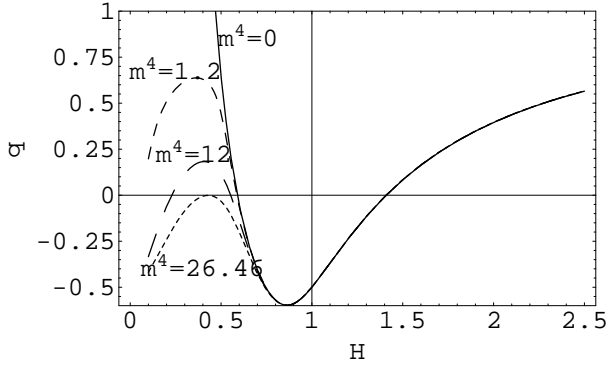}
\caption{ shows the plot of $q$ with $H$ for different values of  $\mu^4$ (shown by $m^4$) with $\alpha=2$ and choose 
the initial conditions as $q$[1]= -0.5, $q'$[1]=1.2.}
\end{figure} 

{$ \bf Special\; Case\;:$}
We consider here  $\alpha =0$ i.e., $f(R)= R -\frac{\mu^4}{R}$. Equation (12) now takes the form 
\begin{equation}
q''- \frac{2q+4}{q^2-1}\; {q'}^2 - \frac{(4q+7)}{(q+1)H}\; q' - \frac{3(q-1)(2q+1)}{(q+1)H^2}  +
\frac{36 (q-1)^3 H^2}{\mu^4 (q+1)} =0 .
\end{equation}
This  case was considered earlier by Das {\it et al.} [24]. However, the equation (see eq. (12)  in Ref. [24])
 considered by them is not correct as some of the terms are missing. Consequently the q vrs. H curve shown in fig. 7 
 here  is found to differ significantly allowing more phases of expansion. We note that the universe in the past may have 
 started from a constant decelerating phase transits to an accelerating phase thereafter once again because of sign flip 
 in $q$, the universe may transit to a decelerating phase. It has been shown for $\mu^4 = 0.01$ that in future the 
 universe may attain a constant decelerating phase and thereafter there will be a sign flip of $q$ once again. As $\mu^4$ 
 is increased the corresponding duration of the present accelerating phase increases. For smaller values in $\mu$ the 
 accelerating phase becomes shorter. However, there are epoch when $q = 1$ is attained before and after the present phase 
of acceleration. In all the cases  universe will end up with an acceleration having $q= -0.5$. As $\mu^4$ increases, the time at which the universe transits from deceleration to acceleration occur at an earlier time and the acceleration to deceleration transition occure at late time. The universe once again transits from deceleration to acceleration and this will occure earlier as $\mu^4$ is increased.

\input{epsf}
\begin{figure}
\epsffile{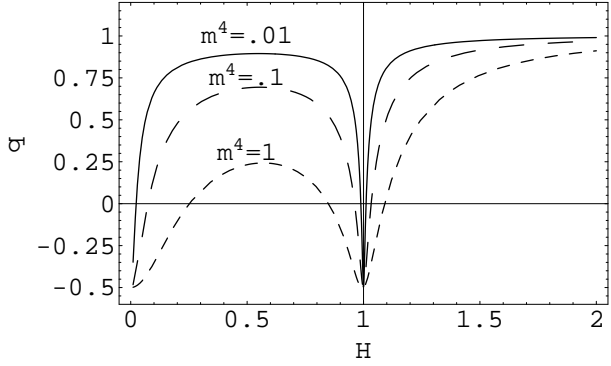}
\caption{shows the plot of $q$ versus $H$ for $f(R) = R -\frac{\mu^4}{R} $ for different value of $\mu^4$ (shown by $m^4$).
 Here we choose the initial conditions as $q$[1]=-0.5, $q'$[1]=1.2.}
\end{figure}

$\bf Case\; B \; :$
 
In this case we  consider  higher derivative theory much discussed in recent times which has the form namely,
  $f(R) = R + b\;ln(R)$,  to look for a physically relevant cosmological model. The relevant differential equation obtained
 from the field equation is given by  
\begin{equation}
q''- \frac{q+3}{q^2-1}\; {q'}^2 - \frac{3}{(q+1)H}\; q' +\frac{2(q-1)^2}{(q+1)} \left[\frac{6}{b}-\frac{1}{H^2}\right] =0 .
\end{equation}
The above equation is  highly nonlinear, consequently  we adopt numerical technique to study the evolution of the universe. 
 Here we look for  the evolution for different values of the coupling parameter $b$ in the action. The curves in fig. 8
 are plotted for different coupling constants $b$ which are interesting.  We note that the universe transits from 
decelerating phase to an accelerating phase in the recent past which will  enter into decelerating phase once again 
in future. For $b < 0$ the duration  for accelerating phase is found to be shorter than that for $b > 0$. 
However, for $b$ positive the duration of the accelerating phase will be longer if $b$ is smaller. 
In the case of negative $b$, the universe is found to land up at a maximum possible acceleration at the present epoch,
 thereafter the rate of acceleration will decrease and consequently transits to decelerating phase once again.
 For $b > 0$, the maximum rate of expansion will be achieved in near future.

\input{epsf}
\begin{figure}
\epsffile{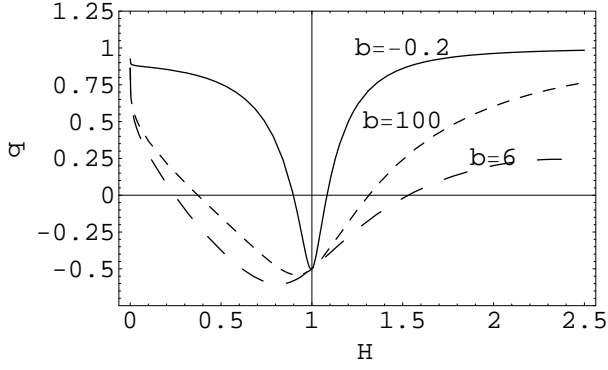}
\caption{shows the plot of $q$ versus $H$ for $f(R) = R + b\;ln{R} $ for different value of $ b $ with $q$[1]=-0.5, $q'$[1]=1.}
\end{figure}

$\bf Case \; C \;:$

In this case we consider  gravitational action with   $f(R) = R + m\; e^{[-nR]}$. 
Consequently, using eq. (8), we obtain the folowing differential equation 
\[
q''+ \frac{1-6n(q+1)H^2}{q+1}\; {q'}^2 + \frac{8q+5 - 24 n(q^2-1) H^2}{(q+1)H}\; q' +
\]
\begin{equation}
 \frac{2(q-1)(3q+4)}{(q+1)H^2} - 24 n (q-1)^2 - \frac{1}{3n^2 m(q+1)H^4}\left[e^{nR}-n\;m\right] =0 .
\end{equation}
The equation is highly non-linear, we adopt numerical technique to solve it as was done in  earlier sections. 
We note the following :

(i)  The variation of $q$ with $H$ for  for different values of $ n $ is shown in the fig.9.
 It is evident that the time of transition of the universe from decelerating to accelerating  phase  depends on $n$.As $n$ is increased, the time at which universe transits from decelerating to accelerating phase is nearer to the present epoch. Also, the duration of the present accelerating phase decreases as $n$ decreases. As $n$ increases, the duration of deceleration to once again deceleration increases.

\input{epsf}
\begin{figure}
\epsffile{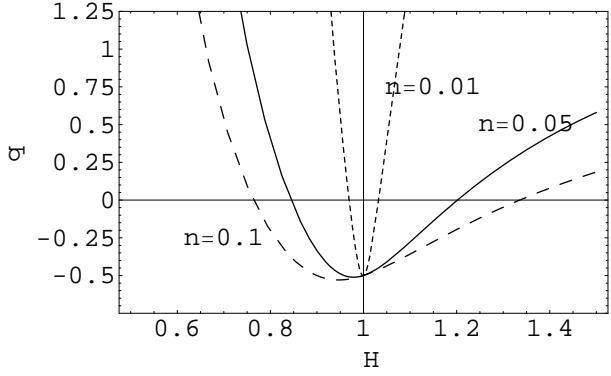}
\caption{ Here we choose the initial conditions as $q$[1]=-0.5, $q'$[1]=1 and $m =4$.}
\end{figure}

(ii)  We plot  $q$ vrs. $H$ for different values of $ m $ taking  $n = 0.1$ which is shown in  fig. 10.
  The curves show that  the universe at the present epoch entered from decelerating to acceleration and it will switch over to decelerating phase once again in future. The smaller
values of $m$ leads to a shorter duration of the present accelerating phase.

\input{epsf}
\begin{figure}
\epsffile{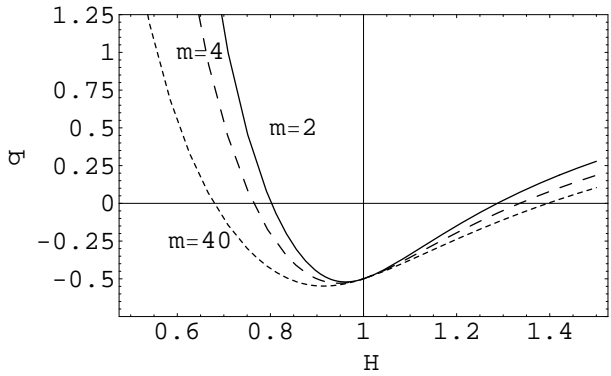}
\caption{shows the plot of $q$ versus $H$ 
for $f(R) = R + m\;Exp[-nR] $ for different value of $m$ with $q$[1]=-0.5, $q'$[1]=1 and $ n=0.1$.}
\end{figure}

From ($ q -  H$) curve as  drawn in fig. 4, we obtain sufficient data set from the numerical plot using {\it Mathematica} [27]. 
Those numerical 
values 
may be used to find a closed analytic mathematical structure for $q$ and $H$ which can be determined using a  polynomial
 function given by $ q = \Sigma _{0}^{n} a_i H^i$.
We now explore for the mathematical function for  $q(H)$ corresponding to fig. 4 with initial value $q'[1]=1.2 $. The 
corresponding approximate analytic function may be expressed as  
\[
q[H]= 45.06381 - 515.24544 H + 2478.58763 H^2 - 6555.42602 H^3 + 10556.26293 H^4\]
\begin{equation}
 -  10835.40545 H^5 + 7153.27419 H^6 - 2949.71637 H^7 + 693.08477 H^8 - 
  70.97952 H^9. 
\end{equation}

\input{epsf}
\begin{figure}
\epsffile{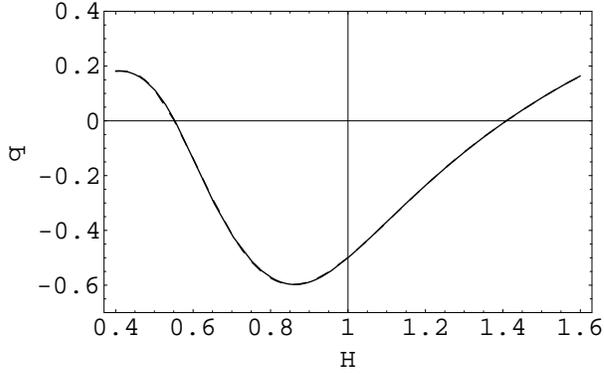}
\caption{ Dashing curve  is due to polynomial expression and Line curve is from fig. 4 with $q'[1] =1.2 $.}
\end{figure} 

We compare the curves obtained numerically  with that corresponding to the curve fitted to an analytic 
 function given in eq. (16). The two curves are shown in  fig.11 which is found exactly superimposed. 
This comparison holds good only when $H$ is reasonably close to one. Thus we  outline how one can determine the 
 approximate relation of $q$ with $H$, which may be employed for other curves also.\\

{\bf IV. DISCUSSION :}

In this paper we obtain cosmological solutions in higher derivative theories of gravity. We explore different phases of expansion that are permitted in the higher derivative theories of gravity without matter as a toy model. Since the field equations obtained from the action  are highly non-linear it is not simple to obtain analytic solution in a closed form. We adopt numerical technique to understand the present and future evolution of the universe here. The relevant field equations corresponding to modified gravitational theories are rewritten in terms of two parameters (i) Hubble parameter (H) and 
(ii)  deceleration   parameter (q) and its derivatives. We plot q vs. H for different values of the coupling constants of the higher derivative gravitational actions namely, (A) $\alpha$, $\mu$; (B)  $b$ and (C) $m$ and $n$ in section III to investigate present and future evolution of the universe.
It is found that all the modified theories of gravity lead to a scenario which accommodate the  present accelerating phase where the universe transits from a decelerating phase. In some cosmological models we  note that the universe might once again transit from the present accelerating phase  to another decelerating phase.     In these models one can determine the  duration of the present accelerating phase in terms of coupling
  parameters in the gravitational action. In figs. (1-10), it is evident that the present accelerating phase of the universe  once
   again will change into a  decelerating phase. In sections fig. (5-7), new cases are observed where yhe  universe from the present accelerating phase transits to a decelerating phase and then the decelerating phase again changes to an accelerating phase which finally ends up at $q=-1/2$ independent of the initial values of the parameter $\mu^4$. Another interesting case is  evident in figs. (1, 3, 4) where the present rate of acceleration of the universe will decrease and attains zero value, where the universe  will grow linearly with time as $a(t) \sim t$  for a sufficient time duration. Thereafter the universe may  once again  enter  into an accelerating phase with an increasing rate of expansion.
A detailed study with matter in the modified gravity will be taken up elsewhere.
       
$\it Acknowledgement$

B. C. P. would like to thank Inter-University centre for Astronomy and Astrophysics (IUCAA), Pune for supporting a visit in which a part of this work was completed. BCP would like to acknowledge the financial support from University Grants Commission, New Delhi through the Minor research Project [Grant No. 32-63/2006 (SR)]. SG would like to thank University of North Bengal for Fellowship.


\begin{thebibliography}{99}
\bibitem{kn:27} MATHEMATICA 3.0 is used to find numerical solution and plotting curves.
\end{thebibliography}
\end{document}